\documentclass[fleqn,twoside]{article}
\usepackage{espcrc2}

\usepackage{epsbox}

\usepackage{graphicx}
\usepackage[figuresright]{rotating}

\newcommand{\AmS}{{\protect\the\textfont2
  A\kern-.1667em\lower.5ex\hbox{M}\kern-.125emS}}

\title{Chiral symmetry on a lattice with hopping interactions}

\author{Takanori Sugihara\address[MCSD]{RIKEN BNL Research Center, 
Brookhaven National Laboratory, Upton, New York 11973, USA}
\thanks{sugihara@bnl.gov}}
       
\begin{document}

\begin{abstract}
The species doubling problem of the lattice fermion is 
resolved by introducing hopping interactions that 
mix left- and right-handed fermions around the momentum boundary. 
Approximate chiral symmetry is realized on the lattice. 
The deviation of the fermion propagator from 
the continuum one is small. 
\vspace{1pc}
\end{abstract}

\maketitle

\section{INTRODUCTION}

In contrast with the great success of lattice gauge theory, 
lattice fermions remain a long-standing problem. 
Naive discretization causes 
the species doubling problem \cite{wilson}. 
The situation does not change 
regardless of how the lattice spacing is reduced 
as long as the space-time derivative is modeled as 
a naive difference. 
Many attempts have been made to fix the doubling problem 
\cite{wilson,ks,slac,kaplan,ch,neuberger,luscher}. 
Wilson removed doublers at low energy by introducing an interaction 
that mixes left- and right-handed fermions \cite{wilson}. 
However, unwanted degeneracy persist at high energy 
and chiral symmetry is explicitly broken. 
To fix these problems, Kaplan modified Wilson's fermion 
with an extra dimension \cite{kaplan} and 
succeeded in realizing approximate chiral symmetry. 
However, the cost of numerical calculations 
based on it is not cheap. 
If we find a method to perform such calculations 
without the extra dimension, 
calculation time decreases largely and a deeper 
understanding of quantum field theory becomes possible. 

The lattice fermion has another serious problem. 
The fermion propagator defined on a lattice deviates 
from the continuum one 
even if the doublers are removed with the existing 
techniques such as Kaplan's fermion \cite{ch}. 
We need to modify the discretized propagator somehow 
so that it is close to the continuum one as far as possible. 

In general, the extra dimension can be expressed as 
hopping interactions in a lower-dimensional system. 
Kaplan's fermion is a formulation with an extra dimension, 
so there must be a corresponding Hamiltonian 
with no extra dimension. 
Also, the shape of the fermion propagator 
can be improved with hopping interactions. 

In this talk, based on a Hamiltonian formalism, 
we introduce ultralocal hopping interactions to remove doublers 
and improve momentum dependence of fermion energy. 
(The word ``ultralocal" means that fermion hopping is 
restricted to a finite range on a real-space lattice 
\cite{bietenholz}.) 
From knowledge of the continuum theory, 
we know the correct momentum dependence of the energy. 
We start from momentum space and go back to real space 
by way of discrete Fourier transform. 
See Ref. \cite{sugi} for the details of this work.

\section{FORMULATION}

Let us consider a free Dirac fermion 
on a (1+1)-dimensional Hamiltonian lattice. 
\begin{equation}
 H=\frac{1}{a}\sum_{l=-N/2+1}^{N/2}
  p_l \bar{\zeta}_l \gamma^1 \zeta_l, 
 \label{h1}
\end{equation}
where $l$ is an index for momentum and $N$ is the number of sites. 
$\zeta_l$ and $\bar{\zeta}_l$ are discrete Fourier transform 
of real-space two-component fermion operators $\psi_n$ 
and $\bar{\psi}_n\equiv\psi_n^\dagger\gamma^0$ 
for $n=1,2,\dots,N$, respectively. 
The system is quantized with anticommutators in the usual way. 
Periodic boundary conditions are assumed in real space, 
$p_l \equiv 2\pi l/N$. 
We try to create a real-space Hamiltonian 
with no doubler that reproduces Eq. (\ref{h1}). 
To find real-space representation of $p_l$, 
let us consider the following function 
\begin{equation}
 \sum_{\alpha=1}^{M}
  \frac{2(-1)^{\alpha-1}}{\alpha}
  \sin \alpha p. 
 \label{sp}
\end{equation}
In the limit $M\to\infty$, the function (\ref{sp}) goes to $p$. 
The function (\ref{sp}) necessarily has a node 
at the boundary $p=\pm\pi$ because it has a periodicity of $2\pi$. 
The node is the cause of the doubling problem. 
The doubler remains as a singularity at the boundary 
even if the limit $M\to\infty$ is taken 
(by ``doubler" we mean unwanted energy degeneracy 
that is not contained in the continuum theory). 
Anyway, the parameter $M$ needs to be small 
for practical formulation 
because $M$ corresponds to the maximum distance 
of fermion hopping. 

\begin{figure}[h]
  \begin{center}
    \epsfile{file=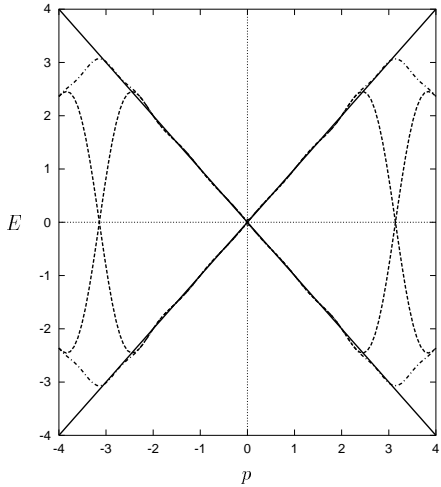,scale=0.7}
  \end{center}
  \caption{The solid line plots the correct energy $\pm p$ 
from the continuum theory. 
The dashed line plots the function $\pm s(p)$ for $M=5$ 
with the Lanczos factor. 
The dot-dashed line plots the function $\pm k(p)$ for 
$M=5$, $u=130$, and $v=8.4$ with the Lanczos factor. 
The function $\pm k(p)$ almost agrees with $\pm p$ 
in the fundamental region $|p|\le\pi$ 
except for a small deviation around momentum $|p|\sim 2.3$. 
}
\label{fig3}
\end{figure}

In addition to the doubler around the momentum boundary, 
the function (\ref{sp}) has another degeneracy. 
It oscillates around $p$ and has local minima 
if the summation is truncated with a small $M$. 
In Fourier analysis, it is called the Gibbs phenomenon, 
which  occurs 
if a function to be expanded has a singularity \cite{aw}. 
The oscillation can be removed by replacing Eq. (\ref{sp}) with 
\begin{equation}
 s(p)=\sum_{\alpha=1}^{M}
   S_\alpha \sin \alpha p, 
 \label{spl}
\end{equation}
where
\[
 S_\alpha \equiv F_\alpha
  \frac{2(-1)^{\alpha-1}}{\alpha}, 
 \quad
 F_\alpha \equiv \frac{M+1}{\pi\alpha}
   \sin\left(\frac{\pi\alpha}{M+1}\right). 
\]
$F_\alpha$ is called the Lanczos factor \cite{aw}. 
The factor almost removes the oscillation of Eq. (\ref{sp}). 
However, the doubler modes around the boundary still remain. 
We are going to remove them 
by a trick with hopping interactions. 
Let us consider the following momentum-space Hamiltonian: 
\begin{equation}
 H=\sum_{l=-N/2+1}^{N/2}
  \left(
   s_l \bar{\zeta}_l \gamma^1 \zeta_l
   +m\bar{\zeta}_l\zeta_l 
  \right), 
 \label{h2}
\end{equation}
where $s_l \equiv s(p_l)/a$ and $m$ is fermion mass. 
We introduce interactions $c_l$ that mix 
left- and right-handed fermions 
\begin{equation}
 H=\sum_{l=-N/2+1}^{N/2} \zeta_l^\dagger
 \pmatrix{
 s_l & m+c_l \cr
 m+c_l & -s_l}\zeta_l, 
 \label{h4}
\end{equation}
where $c_l$ are assumed to be nonzero only for $|l|\sim N/2$. 
The Hamiltonian (\ref{h4}) can be diagonalized. 
\begin{equation}
 H=\sum_{l=-N/2+1}^{N/2} \zeta_l'^\dagger
 \pmatrix{
 k_l & 0 \cr
 0 & -k_l}\zeta_l', 
 \label{h5}
\end{equation}
where $k_l\equiv \sqrt{s_l^2+(m+c_l)^2}$ are 
energies of one particle states 
and $\zeta_l'$ are transformed variables. 
The coefficients $c_l$ are determined 
so that energy of one-particle states 
becomes close to the continuum one as far as possible. 

In Fig. \ref{fig3}, the functions $\pm p$, $\pm s(p)$, 
and $\pm k(p)$ are compared, 
where $k(p)\equiv\sqrt{s^2(p)+c^2(p)}$. 
The function $\pm k(p)$ corresponds to energy of one-particle 
states given by the Hamiltonian (\ref{h4}) with $m=0$. 
The function $\pm k(p)$ agrees well with the correct 
energy $\pm p$ from the continuum theory 
in the fundamental region $|p|\le\pi$. 
The Hamiltonian (\ref{h4}) approximately reproduces 
the continuum theory without doubler 
if we identify $c_l = c(p_l)/a$ and $k_l=k(p_l)/a$. 

\begin{figure}[h]
  \begin{center}
    \epsfile{file=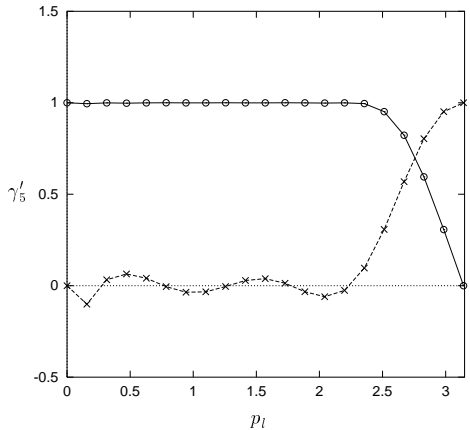,scale=0.7}
  \end{center}
  \caption{
The matrix elements of $\gamma_5'$ are plotted as functions 
of the momentum $p_l$ for $a=1$, $M=5$, $u=130$, and $v=8.4$ 
with the Lanczos factor. 
For a lattice size $N=40$, the functions are plotted 
in the right-half plane of momentum space $p_l\ge 0$. 
The circles are plots of the diagonal (1,1) element 
of $\gamma_5'$. The crosses are plots of minus 
of the off-diagonal (1,2) element of $\gamma_5'$. 
The solid and dashed lines are plotted to guide the eyes. 
}
\label{fig4}
\end{figure}

Figure \ref{fig4} shows the diagonal and off-diagonal 
matrix elements of the transformed $\gamma_5$ 
in the new basis 
that diagonalizes the Hamiltonian (\ref{h4}) with $m=0$. 
The diagonal (1,1) element of $\gamma_5'$ is 
almost unity at low and intermediate energy $p_l<2.3$ 
and deviates from unity at $p_l>2.3$. 
The off-diagonal (1,2) element of $\gamma_5'$ oscillates 
around zero at $p_l<2.3$ and becomes unity at $p_l=\pi$. 
At low energy, the deviation of the off-diagonal elements 
from zero is not large and becomes smaller 
as the parameter $M$ increases. 
At $p_l<2.3$, the transformed left- and right-handed fermions 
have approximately the correct chiral charges $1$ and $-1$, 
respectively. 
The low-energy Hamiltonian (\ref{h4}) has approximate 
chiral symmetry because the commutation relation between 
the Hamiltonian (\ref{h4}) and chiral charge defined 
with $\gamma_5'$ is almost zero for small $l$. 
The errors associated with chiral symmetry can be improved 
in a systematic way by increasing $M$. 
The value used here for the parameter $M$ is sufficiently small 
and does not deny application of the model to 
actual numerical analysis with a computer. 

The real-space Hamiltonian and the Euclidean action 
for the Dirac spinor is obtained 
by substituting the discrete Fourier transform into Eq. (\ref{h4}). 
As usual, gauge symmetry can be implemented by 
inserting exponentiated gauge fields 
between the hopping fermions. 
The continuum limit $a\to 0$ is taken with $M$ fixed. 
The method can be extended to higher-dimensional cases 
by taking care of the doubler of each direction 
in the same way. 

In this talk, I have shown how to 
realize approximate chiral symmetry at low energy. 
Explicit breaking of chiral symmetry has been compressed to 
high energy. 
In future works, it should be precisely checked 
if insertion of gauge interactions affects chiral properties.

\end{document}